\documentclass[aps,pra,epsfigure,twocolumn,showpacs]{revtex4}
\usepackage{amsmath}
\usepackage{amstext}
\usepackage{latexsym}
\usepackage{psfig}
\usepackage{graphicx}
\usepackage{amsfonts}

\newcommand{\ket}[1]{\left\vert#1\right\rangle}
\newcommand{\modul}[1]{\left\vert#1\right\vert}

\newcommand{\one}{\mbox{$1 \hspace{-1.0mm}  {\bf l}$}}

\newcommand{\pro}[2]{\left\vert#1\right\rangle\left\langle#2\right\vert}

\newcommand{\sprod}[2]{\left\langle#1\vert#2\right\rangle}

\newcommand{\bra}[1]{\left\langle#1\right\vert}

\newcommand{\valmed}[1]{\left\langle#1\right\rangle}
\newcommand{\num}[1]{\hat{n}_{#1}}

\begin{document}

\title{Perspectives for quantum state engineering via high non-linearity in a double-EIT regime}
\author{M. Paternostro}
\affiliation{School of Mathematics and Physics, The Queen's University,
Belfast BT7 1NN, United Kingdom}
\author{M. S. Kim}
\affiliation{School of Mathematics and Physics, The Queen's University,
Belfast BT7 1NN, United Kingdom}
\author{B. S. Ham}
\affiliation{Center for Quantum Coherence and Communications, Electronics and Communications Research Institute, Daejeon, 305-350, South Korea}
\date{\today}

\begin{abstract}
We analyse the possibilities for quantum state engineering offered by a model for Kerr-type non-linearity enhanced by electromagnetically induced transparency (EIT), which was recently proposed by Petrosyan and Kurizki [{\sl Phys. Rev. A} {\bf 65}, 33833
(2002)]. We go beyond the semiclassical treatment and 
derive a quantum version of the model with both a full Hamiltonian approach and an analysis in terms of dressed states.  
The preparation of an entangled coherent state via a cross-phase modulation effect is demonstrated. We briefly show that the violation of locality for such an entangled coherent state is robust against low detection efficiency. Finally, we investigate the possibility of a bi-chromatic photon blockade realized via the interaction of a low density beam of atoms with a bi-modal electromagnetic cavity which is externally driven. We show the effectiveness of the blockade effect even when more than a single atom is inside the cavity. The possibility to control two different cavity modes allows some insights into the generation of an entangled state of cavity modes.
\end{abstract}
\maketitle

\section{Introduction}
\label{introduzione}

The reliable preparation of non-classical states of light such as a travelling-wave 
entangled coherent state~\cite{ecs,cohe} and Schr\"odinger cat state~\cite{gatto}, or the control of the population of an individual field mode (of an electromagnetic cavity, for example) are recognized to be important tasks in Quantum Information Processing (QIP)~\cite{NielsenChuang}. Entangled coherent states and Schr\"odinger cats, for example, revealed useful for QIP with coherent states~\cite{cohe}. On the other hand, photon blockade appears as a striking manifestation of control on a system at the quantum level and opens a way to novel schemes for quantum state engineering~\cite{Imamoglublockade}.

In this paper, we investigate the possibilities of quantum engineering using non-linear processes realized by electromagnetically induced transparency (EIT)~\cite{Arimondo}. The EIT regime seems to be able to overcome one of the major bottlenecks in a non-linear process: the low efficiency accompanied by high absorption rate of a conventional Kerr medium, that makes the production of a travelling-wave cat state, for example, far from realization~\cite{YurkeStoler}.

On the other hand, it has been proved that the atomic medium in EIT regime shows a measured $\chi^{(3)}$ parameter up to six orders of magnitude larger
than usual~\cite{Hau}. This suggest the use of the enormous non-linearity to get a reliable {\sl cross phase modulation} effect between two
travelling fields of light even for the very low photon-number case~\cite{SchmidtedImamoglu,HarrisHau}. Usually, the approach to such a process 
is semiclassical. But, for the purposes of QIP, a full quantum treatment is relevant~\cite{opticalFredkingate,cohe}. Such an analysis has been performed in ref.~\cite{LukinImamoglu}, where the idea of a double-EIT regime is introduced in order to optimize the non-linear interaction between 
two electromagnetic fields. The proposed model has been modified, in~\cite{PetrosyanKurizki}, to get an easier experimental 
realization of the process. In this latter work, however, the analysis is again semiclassical. We give the full quantum mechanical description of~\cite{PetrosyanKurizki} by means of an Hamiltonian approach~\cite{Kryzhanovsky}. 

A promising candidate for the embodiment of the atomic model we discuss is a Pr$^{3+}$ doped Y$_{2}$SiO$_{5}$ crystal (Pr:YSO): it has an energy-level scheme appropriate for our purposes and it has been used for the demonstration of EIT~\cite{ham1} and giant non-linearity in solid state devices~\cite{ham2,turukhin}. Using typical values for Pr:YSO, we find a giant non-linearity even at the quantum level. We derive the equations of motion 
for the involved quantum fields from an effective interaction
Hamiltonian. With these results, starting from two independent coherent states, we prepare an entangled coherent state and a Schr\"odinger cat state. 

As a second example of the applicability of these results, we treat a cavity-quantum-electrodynamics (CQED) system. We concentrate on a photon-blockade effect realized combining the obtained large non-linearity and the features of isolation from the environment and manipulability characteristic of CQED. As we see, the interaction of the atomic model we depict with the electromagnetic field of a cavity results in a coupled system that exhibits a non-linear eigenspectrum~\cite{Imamoglublockade}. This is expoitable to control the number of excitations that are fed into the cavity from an outside radiation. Furthermore, the specific model for a double-EIT allows to treat the interaction with a bi-chromatic cavity field and to show how to manipulate it to settle entanglement.

The paper is organized as follows: in Section~\ref{ilmetodo} we sketch the Hamiltonian method we have chosen; it
is immediatly applied, in Section~\ref{doubleEIT}, to the atomic model for double-EIT~\cite{PetrosyanKurizki}.
We then derive the equations of motion for the quantized fields and give the order of magnitude of the achieved rate of non-linearity. Section~\ref{appendice} is devoded to an alternative approach to the double-EIT problem: we choose a dressed state picture to re-derive the polarizabilities of the medium. The effective interaction Hamiltonian derived in Section~\ref{doubleEIT} is used to show, in Section~\ref{catstates}, how a tensorial product of two coherent states evolves toward an entangled coherent state. A technique to project one of the modes onto a Schr\"odinger cat state is analyzed. We describe a scheme for the characterization of a generated cat state~\cite{duan}. In Section~\ref{photonblockade} we give the outlines of a bi-chromatic photon-blockade realized, in an electromagnetic cavity, by the high rate of non-linearity inherent in the chosen model for double-EIT.  

\section{The Hamiltonian method}
\label{ilmetodo}

Usually, the interaction of electromagnetic fields with an atomic medium is mathematically described by means of the Maxwell-Bloch equations, which are a set of coupled differential equations that connect the dynamics of the fields to that of the atomic degrees of freedom. These latter evolve according to the Von Neumann equation $i\hbar\frac{\partial\rho(t)}{\partial{t}}=[H,\rho(t)]$, with $\rho(t)$ the atomic density operator and $H=H_{atom}+H_{field}+H_{interaction}$ the complete Hamiltonian model for the problem.
Whenever conditions of adiabaticity and low intensities of the fields are valid, a steady state solution can be obtained. Inserting it into the Maxwell equations gives the evolution of the fields.

In an EIT problem, however, the number of interacting fields as well as the involved atomic levels is usually large. This makes the analytical solution of the Maxwell-Bloch equations a challenging task. From classical considerations, it is possible to see that the polarization $P_{j}$ of a medium 
can be expressed as:
\begin{equation}
\label{polarization}
P_j=-\frac{Nd_{j}}{\hbar}\left\langle\frac{\partial H'}{\partial \Omega^{*}_j}\right\rangle{e^{-i(\omega_{j}t-k_{j}z)}}+c.c.,
\end{equation}
where $\Omega_{j}$ is the Rabi frequency of the j-th field (of frequency $\omega_{j}$), $d_{j}$ is the dipole matrix element of the corresponding transition, $N$ is the density of the atomic medium and $H'$ is the single-particle interaction Hamiltonian. Here we assume that the atoms in the medium are equally coupled to the different fields. When Eq.~(\ref{polarization}) is introduced into the Maxwell-Bloch equations and we use the slowly varying envelope approximation (SVEA), we get:
\begin{equation}
\label{mb}
\left(\frac{\partial}{\partial{z}}+\frac{1}{c}\frac{\partial}{\partial{t}}\right)\Omega_{j}=-i\frac{Nd^{2}_{j}\omega_{j}}{2\hbar{\epsilon_0}c}\left\langle\frac{\partial H'}{\partial \Omega^*_{j}}\right\rangle,\hskip0.3cm\forall{j}.
\end{equation}

The fields responsible for the coupling of the initially prepared atomic state to 
other levels are taken, here, to be weak (weak coupling limit). This gives a small probability of transition toward states different from the initial one. The initial state becomes, thus, a kind of stationary state, whose evolution will be adiabatically followed by the ensemble. In Eq.~(\ref{mb}), $H'$ can thus be replaced by $\lambda$, the energy eigenvalue of the initially prepared state.

The quantization of the interacting fields can be done giving operatorial nature to the field variables in the effective Hamiltonian that $\lambda$ represents. Bosonic commutation rules to the fields creation and annihilation operators are thus imposed~\cite{landau}. 


\section{Cross phase modulation via a double EIT effect}
\label{doubleEIT}

\begin{figure}[ht]
\centerline{\psfig{figure=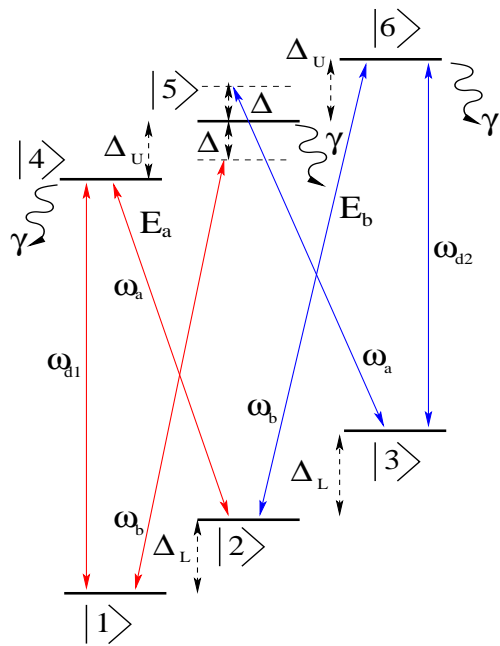,width=6.5cm,height=8.5cm}}
\caption{The atomic model for double-EIT. Fields $E_a$ and
$E_b$ have frequency $\omega_a$ and
$\omega_b$, respectively. The fields with frequencies
$\omega_{d1},\hskip0.1cm\omega_{d2}$ are classical (their intensities are much greater than 
those of $E_a$ and $E_b$). Detunings $\Delta_U\neq\Delta_L$ and $\modul{\Delta}=\modul{\Delta_U-\Delta_L}$. 
The decay rates of the excited states are taken equal to $\gamma$, for the sake of simplicity.} 
\label{Petrosyan}
\end{figure}

Here we describe the model for double-EIT proposed in~\cite{PetrosyanKurizki} (see Fig.~\ref{Petrosyan}). 

It involves a ground metastable triplet $\left\{\ket{1},\ket{2},\ket{3}\right\}$ and an excited one $\left\{\ket{4},\ket{5},\ket{6}\right\}$. By means of an external magnetic field the ground states are splitted by
$\Delta_{L}$ while $\Delta_{U}$ $(\neq\Delta_{L})$ is the splitting for the excited triplet. Two weak fields, $E_a$ and $E_b$, drive resonantly $\ket{2}\leftrightarrow\ket{4}$ and $\ket{2}\leftrightarrow\ket{6}$ respectively; $\ket{1}\leftrightarrow\ket{5}$ and $\ket{3}\leftrightarrow\ket{5}$ are in the dispersive regime (detuning $\modul{\Delta}=\modul{\Delta_{U}-\Delta_{L}}$) while transition $\ket{2}\leftrightarrow\ket{5}$ is assumed to be forbidden. The couplings $\ket{1}\leftrightarrow\ket{4}$ and
$\ket{3}\leftrightarrow\ket{6}$ are realized by two classical fields of different frequencies but equal Rabi
frequencies. Two distinct subsystems are easily singled out: states $\left\{\ket{1},\ket{4},\ket{2},\ket{5}\right\}$ constitute the four-level ${\sf N}$  system proposed in~\cite{SchmidtedImamoglu} to show giant Kerr non-linearity. EIT is realized for field $E_a$ while $E_{b}$ acts as a perturbation that induces a.c. Stark shift on $\ket{1}$, which determines a large non-linear effect~\cite{LukinImamoglu}. An
analogous description can be done, interchanging $E_b$ with $E_a$, for
the subset $\{\ket{3},\ket{6},\ket{2},\ket{5}\}$. The two subsystems are connected by the non-resonant couplings
to $\ket{5}$. In the interaction picture, the Hamiltonian reads:
\begin{equation}
\label{Hamiltoniana}
\begin{aligned}
H'&=\hbar\left\{\{\Delta\pro{5}{5}+\Omega_{d}\pro{4}{1}+\Omega_{d}\pro{6}{3}+\Omega_{a}\pro{4}{2}\right.\\
&\left.+\Omega_{b}\pro{5}{1}+\Omega_{b}\pro{6}{2}+\Omega_{a}e^{-2i\Delta{t}}\pro{5}{3}+c.c.\right\}.
\end{aligned}
\end{equation}

We then introduce the decay rates of the excited states $\gamma$ and {\sl phenomenologically} change the signs in front of each Rabi frequency in Eq.~(\ref{Hamiltoniana}) to match the expression in ref.~\cite{PetrosyanKurizki}. 
The atoms in the ensemble can be prepared in state $\ket{2}$ by means of optical pumping and, if the weak field limit is assumed ($\modul{\Omega_d}\gg\modul{\Omega_{a,b}}$), the atomic system remains in $\ket{2}$ all along the interaction time.

To discard the Doppler broadening, an atomic gas can be mantained at a low temperature. Alternatively, we can take a solid state device as the Pr$^{3+}$ doped Y$_{2}$SiO$_{5}$ crystal (Pr:YSO)~\cite{equall} to embody the Hamiltonian model. This latter choice is motivated by the similarity between the energy-level scheme described here and that of the transition ${^3H}_{4}\rightarrow{^1D}_{2}$ in Pr:YSO. In~\cite{turukhin}, ultraslow group velocity ($\simeq45$ m/sec) in Pr:YSO has been reported. 

Following the recipe outlined in Section~\ref{ilmetodo}, we solve the secular equation for $H'$ and seek for the eigenvalue of the initial state $\ket{2}$. In the weak field limit we get:
\begin{equation}
\label{soluzionemiaapprox}
\lambda\simeq\frac{2\hbar\modul{\Omega_{a}}^2\modul{\Omega_{b}}^2}{(i\gamma-\Delta)\modul{\Omega_{d}}^2}.
\end{equation}
The derivative of Eq.~(\ref{soluzionemiaapprox}) with respect to $\Omega^*_{a}$ ($\Omega^*_{b}$) gives the polarization of the medium at frequency $\omega_{a}$ ($\omega_{b}$). The equation of motion for $\Omega_{a}$ can be, finally, cast into:
\begin{equation}
\label{moto}
\left(\frac{\partial}{\partial{x}}+\frac{1}{c}\frac{\partial}{\partial{t}}\right)\Omega_{a}=\frac{2iN\sigma_{0}\gamma\modul{\Omega_{b}}^2}{(\gamma+i\Delta)\modul{\Omega_{d}}^2}\Omega_{a}=i\alpha_{a}\Omega_{a},
\end{equation}
with $\alpha_{a}$ the atomic polarizability at frequency $\omega_{a}$~\cite{PetrosyanKurizki}. Here, $\sigma_{0}=\frac{\modul{d}^{2}\omega}{2\epsilon_{0}c\hbar\gamma}$ is the resonant absorption cross section. The equation for $E_{b}$ can be analogously derived.

It has been shown in~\cite{HarrisHau} that the interaction between two fields in a medium exhibiting EIT is 
critically limited by the time that the faster of them spends inside the medium itself. This velocity mismatch strongly affects the efficiency of any
non-linear process we want to realize in the usual {\sf N} configuration. In~\cite{LukinImamoglu}, the induction of an EIT 
regime for both the fields ({\sl double}-EIT) is suggested to bypass the problem. Strongly reducing the group velocities of the beams (by means of EIT), the interaction time could be maximized, optimizing the efficiency of the non-linear process. If we compute the group velocities for fields $E_{a}$ and $E_{b}$ in the model described above, we find $v^{group}_{a,b}\simeq{\modul{\Omega_{d}}^2}/{N\sigma_{0}\gamma}\ll{c}$. This is the signature of the double-EIT established in the atomic ensemble. 

We now demonstrate how this effect is useful for a strong effective non-linear effect on the evolution of $E_{a}$ and $E_{b}$ in the full quantum domain. We give operatorial nature to the Rabi frequencies in Eq.~(\ref{soluzionemiaapprox}) introducing the positive and negative
frequency components of the corresponding operators. For example: $\hat{\Omega}_{a}(z,t)=d_{24}\sum_{k}\sqrt{\frac{\omega^{car}_{a}}{2\hbar\epsilon_{0}V_{q}}}\hat{a}_{k}(t)e^{-i(\omega_{k}-\omega^{car}_{a})z/c}$ (and analogous for $\hat{\Omega}_{b}$). This expression well describes a pulse in the narrow bandwidth approximation: $k$ is a label for the wavelengths in the packet, $\omega^{car}_{a}$ is the central ({\it carrier}) frequency of the pulse and $V_{q}$ is the quantization volume. The narrow bandwidth approximation takes $\delta\omega\ll\omega^{car}_{a}$, where $\delta\omega$ is the bandwidth of the pulse. These operators satisfy the commutation rules $[\hat{\Omega}_{i},\hat{\Omega}^{\dagger}_{j}]\propto\delta_{ij}\hat{\one}$ ($i,j=a,b$),
with $\delta_{ij}$ the Kronecker symbol and $\hat{\one}$ the identity operator. Multiplying by the atomic density $N$ and integrating over the interaction volume $V$, we have the effective Hamiltonian:
\begin{equation}
\label{Hamiltonianeffettiva}
\hat{H}_{eff}=\frac{2\hbar{N}}{(i\gamma-\Delta)}\int_{V}
\frac{\hat{\Omega}^{\dagger}_{a}\hat{\Omega}_{a}\hat{\Omega}^{\dagger}_{b}\hat{\Omega}_{b}}
{\modul{\Omega_{d}}^2}dV.
\end{equation}

We can now write the Heisenberg equations for $\hat{E}_{a}$ and $\hat{E}_{b}$. The evolution of the probe fields is, then, given by:
\begin{equation}
\label{pulsespropagation}
\hat{E}_{a,b}(L,t)=\hat{E}_{a,b}(0,t')e^{i\tilde\chi\hat{E}_{b,a}^{\dagger}(0,t')\hat{E}_{b,a}(0,t')}
\end{equation}  
with $t'=t-L/v^{group}$, ${\tilde\chi}$ a non-linearity rate (obtained collecting all the non operatorial quantities in Eq.~(\ref{Hamiltonianeffettiva})) and $L$ the interaction length. The above equation shows explicitly an effect of cross phase modulation on the quantum fields. For a cw laser beam, we can recast Eq.~(\ref{Hamiltonianeffettiva}) as:
\begin{equation}
\label{chialto}
\begin{aligned}
&\hat{H}^{cw}_{eff}=\hbar\chi\hat{a}^{\dagger}\hat{a}\hat{b}^{\dagger}\hat{b}\\
\text{with}\hskip0.4cm\chi=&\Re\left\{\frac{N\omega_{a}\omega_{b}\modul{d_{24}}^2\modul{d_{26}}^2}{2\hbar^2\epsilon_{0}^2(i\gamma-\Delta)\modul{\Omega_{d}}^2V}\right\},
\end{aligned}
\end{equation}
where we have assumed $V\equiv{V}_{q}$ and $\Re\{\cdot\}$ is to take the real part. We can write and solve the Heisenberg equations for $\hat{a}$ and $\hat{b}$, representative of the dynamics of $E_{a}$ and $E_{b}$ respectively, getting:
\begin{equation}
\label{cw}
\hat{a}_{out}(t)=e^{-i\chi{t}\num{b}}\hat{a}(0)\hskip1cm\hat{b}_{out}(t)=e^{-i\chi{t}\num{a}}\hat{b}(0),
\end{equation}
where $\hat{n}_{j}$ is the photon number operator for field $j=a,b$.
 
The evolution of one field rises up depending, in the most explicit way, on the intensity of the other one. To estimate the order of magnitude of $\chi$ we use values for the parameters in~(\ref{chialto}) typical of the ${^3H}_{4}\rightarrow{^1D}_{2}$ transition in Pr:YSO. We take a wavelength of $\sim600$ nm~\cite{turukhin}, $L\sim{1}$ mm and a cross section for the beams of $100$ $\mu$m (Full-Width-at-Half-Maximum). The decay rate $\gamma$ can be taken between $10$ and $100$ kHz (sample's temperature of $5$ K); $\Delta\sim1$ MHz and $\modul{\Omega_{d}}\sim1$ MHz are reasonable values and allow us to consider $\gamma\ll\Delta,\hskip0.1cm\modul{\Omega_{d}}$, which gives a negligible rate of two-photon-absorption. Indeed, two-photon absorption appears because of the two-photon resonant driving at the basis of EIT. The absorption is proportional to the imaginary part of the polarizability of the medium and it can be minimized if $\Delta\gg\gamma$. The electric dipole matrix elements for the system in exam are typically $\sim10^{-32}$ Cm and $N\sim10^{15}$ cm$^{-3}$. With these values in Eq.~(\ref{chialto}) and for interaction times $T\sim\mu$sec, a cross-phase shift $\chi{T}=\pi$ is achieved, even with beam intensities of just a few of photons. 

An unwanted effect comes from the, here neglected, couplings of $E_{a}$, $E_{b}$ with the atoms via the transitions $\ket{1}\leftrightarrow\ket{5}$ and $\ket{3}\leftrightarrow\ket{5}$ respectively. These spurious couplings lead to the {\it self-phase modulation}, an effect for which a field evolves independently from the other one. The polarizability, in this case, scales as $\alpha^{self}_{j}\propto\Omega^2_{j}/(\Delta_{U}+\Delta_{L})$. Because this is smaller than the effect of cross phase modulation and independent from it, we can safely neglect it when dealing with cross effects. 
 

\section{Atomic polarizabilities in a dressed states picture}
\label{appendice}

In what follows, we change the point of view and we treat the problem of double-EIT using a Hamiltonian model in which the fields are quantized {\it ab initio}. We refer again to Eq.~(\ref{Hamiltoniana}) but we now make the subtitutions $\Omega_{a}\pro{4}{2}\rightarrow{g}_{a}\hat{a}\pro{4}{2}$, $\Omega_{b}\pro{6}{2}\rightarrow{g}_{b}\hat{b}\pro{6}{2}$, where $g_{a,b}$ are related to each coupling. Clearly, we have to add, to Eq.~(\ref{Hamiltoniana}), the term $\hbar(\omega_{a}\hat{n}_{a}+\omega_{b}\hat{n}_{b})$ that is the energy of the quantized fields. We, thus, can write:
\begin{equation}
\label{Hamiltonianaquantizzata}
\begin{aligned}
H'=&\hbar\left\{\Omega_{d}\pro{4}{1}-\Omega_{d}\pro{6}{3}\right.\\
&\left.+g_{a}\hat{a}\pro{4}{2}-g_{b}\hat{b}\pro{6}{2}+c.c.\right\}
\end{aligned}
\end{equation}

where the couplings $\ket{1}\leftrightarrow\ket{5}$ and $\ket{3}\leftrightarrow\ket{5}$ have not been introduced, yet. We justify this choice later in this Section. It is possible to show that the states coupled by the Hamiltonian (\ref{Hamiltonianaquantizzata}) are the elements of the following set:
\begin{equation}
\nonumber 
\begin{aligned}
\{\ket{1,n_{a}-1,n_{b}},&\ket{2,n_{a},n_{b}},\ket{3,n_{a},n_{b}-1},\\
&\ket{4,n_{a}-1,n_{b}},\ket{6,n_{a},n_{b}-1}\}. 
\end{aligned}
\end{equation}
The matrix representation for $H'$, in the Hilbert space restricted to these states, is:
\begin{equation}
\label{Hamiltonianappendix}
H'=\hbar
\begin{pmatrix}
0&0&0&\Omega_{d}&0\\
0&0&0&g_{a}\sqrt{n_{a}}&-g_{b}\sqrt{n_{b}}\\
0&0&0&0&-\Omega_{d}\\
\Omega_{d}&g_{a}\sqrt{n_{a}}&0&0&0\\
0&-g_{b}\sqrt{n_{b}}&-\Omega_{d}&0&0
\end{pmatrix}
\end{equation}
with the Rabi frequencies taken as real. The diagonalization of (\ref{Hamiltonianappendix}) leads to the eigenvalues:
\begin{equation}
\label{eigenvalues}
\begin{aligned}
&{E}^{}_{0}=0\\
{E}^{-}_{1}=-\hbar{\Omega_{}}&\hskip1cm{E}^{+}_{1}=\hbar{\Omega_{}}\\
{E}^{-}_{2}=-\hbar{\Omega_{d}}&\hskip1cm{E}^{+}_{2}=\hbar{\Omega_{d}},
\end{aligned}
\end{equation}
with $\Omega=\sqrt{\Omega_{d}^{2}+g^{2}_{a}(n_{a})+g^{2}_{b}(n_{b})}$. The eigenstates can easily be found and we call them $\left\{\ket{0_{n_{a}n_{b}}},\ket{1^-_{n_{a}n_{b}}},\ket{1^+_{n_{a}n_{b}}},\ket{2^-_{n_{a}n_{b}}},\ket{2^+_{n_{a}n_{b}}}\right\}$. This dressed eigensystem shows that the model can be mapped into a five-level ladder model. This map is just a rotation of the bare basis into the dressed one: $\ket{\mathbf {dressed}}={P}\ket{\mathbf {bare}}$, where ${P}$ is the matrix that realizes the rotation; $\ket{\mathbf {dressed}}$ and $\ket{\mathbf {bare}}$ are two vectors of dressed and bare states, respectively.

On the dressed states, Eq.~(\ref{Hamiltonianaquantizzata}) is diagonal and takes the form:
\begin{equation}
H'_{dressed}=\hbar\sum_{i=1,2}\sum_{j=\pm}E^{j}_{i}\pro{i^{j}_{n_a{n}_{b}}}{i^{j}_{n_a{n}_{b}}}.
\end{equation}
We introduce the couplings to level $\ket{5}$ writing the spin-flip operators $\pro{5}{1}$ and $\pro{5}{3}$ in terms of dressed states. To do it, we use a closure relation so that, for example: 
\begin{equation}
\begin{aligned}
\hat{b}\pro{5}{1}&=\sum_{n_{a},n_{b}}\sum_{n'_{a},n'_{b}}\sqrt{n_{b}+1}\ket{5{n}_{a}n_{b}}\left\{\frac{g_{a}\sqrt{n'_{a}}}{\Omega'}\bra{0_{n'_{a}n'_{b}}}\right.\\
&+\frac{\Omega_{d}g_{a}\sqrt{n'_{a}}}{\sqrt{2}\Omega'\delta'}\bra{1^{-}_{n'_{a}n'_{b}}}-\frac{\Omega_{d}g_{a}\sqrt{n'_{a}}}{\sqrt{2}\Omega'\delta'}\bra{1^{+}_{n'_{a}n'_{b}}}\\
&\left.-\frac{g_{b}\sqrt{n'_{b}}}{\sqrt{2}\delta'}\bra{2^{-}_{n'_{a}n'_{b}}}+\frac{g_{b}\sqrt{n'_{b}}}{\sqrt{2}\delta'}\bra{2^{+}_{n'_{a}n'_{b}}}\right\}
\end{aligned}
\end{equation} 
with $\delta'=\sqrt{g^{2}_{a}n'_{a}+g^{2}_{b}n'_{b}}$. Analogous expressions can be derived for the other field-atom operators. Furthermore, the term $\Delta\pro{5}{5}$ has to be added to~(\ref{Hamiltonianaquantizzata}). This way to introduce the bare atomic level $\ket{5}$ reminds a {\it perturbative} approach justified because of the dispersive nature of the couplings to this level. In the weak field limit and for a sufficiently large detuning $\Delta$, the transition probability to $\ket{5}$ remains small.

To shorten the notation, in the following we take $g_{a}\sqrt{n_{a}}\equiv\tilde{\Omega}_{a}$, $g_{b}\sqrt{n_{b}}\equiv\tilde{\Omega}_{b}$. Applying the Hamiltonian operator to a generic state vector, decomposed as:
\begin{equation}
\begin{aligned}
\ket{\psi}&=A_{0}\ket{0_{n_{a}n_{b}}}+A^{-}_{1}\ket{1^{-}_{n_{a}n_{b}}}+A^{+}_{1}\ket{1^{+}_{n_{a}n_{b}}}\\
&+A^{-}_{2}\ket{2^{-}_{n_{a}n_{b}}}+A^{+}_{2}\ket{2^{+}_{n_{a}n_{b}}}+A_{5}\ket{5n_{a}n_{b}},
\end{aligned}
\end{equation}
gives a Schr\"odinger equation that is equivalent to the following set of differential equations:
\begin{equation}
\label{setappendixl1}
\begin{aligned}
&i\partial_{t}A_{0}=-\frac{\tilde{\Omega}_{a}\tilde{\Omega}_{b}}{\Omega}A_{5}\\
&{i}\partial_{t}A^{-}_{1}=E^{-}_{1}A^{-}_{1}-\frac{\tilde{\Omega}_{a}\tilde{\Omega}_{b}\Omega_{d}}{\sqrt{2}\Omega\delta}A_{5}\\
&{i}\partial_{t}A^{-}_{2}=E^{-}_{2}A^{-}_{2}+\frac{\tilde{\Omega}^{2}_{b}}{\sqrt{2}\delta}A_{5}\\
&i\partial_{t}A^{+}_{1}=E^{+}_{1}A^{+}_{1}+\frac{\tilde{\Omega}_{a}\tilde{\Omega}_{b}\Omega_{d}}{\sqrt{2}\Omega\delta}A_{5}\\
&{i}\partial_{t}A^{+}_{2}=E^{+}_{2}A^{+}_{2}-\frac{\tilde{\Omega}^{2}_{b}}{\sqrt{2}\delta}A_{5}\\
&{i}\partial_{t}A_{5}=\Delta{A}_{5}+\frac{\tilde{\Omega}^{2}_{b}}{\sqrt{2}\delta}\left(A^{+}_{2}-A^{-}_{2}\right)\\
&-\frac{\tilde{\Omega}_{a}\tilde{\Omega}_{b}}{\Omega}A_{0}+\frac{\tilde{\Omega}_{a}\tilde{\Omega}_{b}\Omega_{d}}{\sqrt{2}\Omega\delta}\left(A^{+}_{1}-A^{-}_{1}\right)
\end{aligned}
\end{equation}

We have neglected terms oscillating at frequency $2\Delta+E^{\pm}_{1,2}$ because they average to zero when the time integrals are carried out. Here, again, we adopt SVEA and we assume as initial state $\ket{2}$. The contribution of this bare state to the linear combinations that define the dressed ones is relevant only for $\ket{0_{n_{a}n_{b}}}$, that has the form
\begin{equation}
\label{stato0}
\begin{aligned}
\ket{0_{n_{a}n_{b}}}=&-\frac{\tilde{\Omega}_{a}}{\Omega}\ket{1,n_{a}-1,n_{b}}+\frac{\Omega_{d}}{\Omega}\ket{2,n_{a},n_{b}}\\
&-\frac{\tilde{\Omega}_{b}}{\Omega}\ket{3,n_{a},n_{b}-1},
\end{aligned}
\end{equation}
while it is of order $\delta/\Omega\ll{1}$ or even null for all the other dressed states. Thus, for $\Omega_{d}\simeq\Omega$, we can take $A_{0}=1$. Moreover, it is easy to show that $A^{+}_{1}(t)=A^{-}_{1}(t)$ and $A^{-}_{2}(t)=A^{+}_{2}(t)$ which leads to $A_{5}(t)\simeq\tilde{\Omega}_{a}\tilde{\Omega}_{b}/(\Omega\Delta)$.
We note that $\ket{0_{n_{a}n_{b}}}$ is a {\it dark state} since it is composed of the ground states only and does not contain the decaying states $\ket{4}$ and $\ket{6}$. All the other eigenstates have a contribution from both $\ket{4}$ and $\ket{6}$ and are {\sl bright} ones.
We can derive all the other probability amplitudes:
\begin{equation}
A^{+}_{1}=A^{-}_{1}=\frac{\Omega^{2}_{a}\Omega^{2}_{b}\Omega_{d}}{\sqrt{2}\Omega^{3}\delta\Delta}\hskip0.7cm{A}^{+}_{2}=A^{-}_{2}=\frac{\Omega_{a}\Omega^{3}_{b}}{\sqrt{2}\Omega\delta\Delta\Omega_{d}}.
\end{equation} 

If we want the atomic polarizability at frequency $\omega_{a}$, as in Eq.~(\ref{moto}), we have to come back to the bare state description. To find the probability amplitudes for each bare atomic state, we write $\ket{\psi}=\sum^{6}_{i,1}\beta_{i}\ket{i}$ and equate it to its expression in terms of dressed states. We look for the coefficients $\beta_{2},\beta_{3},\beta_{4},\beta_{5}$, since the polarizability $\alpha_{a}$ is proportional to $(\beta^{*}_{2}\beta_{4}+\beta^{*}_{3}\beta_{5})$. Clearly $\beta_{5}=A_{5}$ and for the others, we have:
\begin{equation}
\begin{aligned}
\beta_{2}=\sprod{2}{\psi}=-\frac{\Omega_{d}}{\Omega}&\hskip1cm\beta_{3}=\sprod{3}{\psi}=\frac{\Omega_{b}}{\Omega}\\
\beta_{4}=\sprod{4}{\psi}=&\frac{\Omega^{2}_{b}\Omega_{a}}{\delta^{2}\Omega\Delta}\left\{\frac{\Omega^{2}_{b}}{\Omega_{d}}-\frac{\Omega^{2}_{a}\Omega_{d}}{\Omega^{2}}\right\}
\end{aligned}
\end{equation}
so that $\alpha_{a}\simeq\frac{N\sigma_{0}\tilde\Omega_{a}\tilde\Omega^{2}_{b}}{\Delta\Omega_{d}^{2}}$, with $\sigma_{0}$ as defined in Section~\ref{doubleEIT}.
This result matches what has been obtained by the Hamiltonian approach once we replace the field variables with the corresponding operators. The dressed states approach described here and the Hamiltonian one lead to consistent results. Since the former relies on an {\it ab initio} quantum level, an undeniable reliability is given to the latter method.
 

\section{Schr\"odinger cat states generation}
\label{catstates}
We apply the results obtained in the full quantized picture of the cross phase modulation via double-EIT to produce non-classical states of field mode. The evolution shown in Eq.~(\ref{cw}) can be derived from the action of the unitary operator $\hat{U}(\varphi(t))=e^{-i\varphi(t)\num{a}\num{b}}$, with $\varphi(t)=\chi{t}$ \cite{Sandersme}. If the initial state of the field modes $E_{a}$, $E_{b}$ is the tensorial product of two coherent states $\ket{\psi(0)}_{ab}=\ket{\alpha}_{a}\otimes\ket{\gamma}_{b}$, its evolution by means of $\hat{U}(\varphi)$, for $\varphi(T)=\pi$, is given by~\cite{Sandersme}
\begin{equation}
\label{quasicatstates}
\ket{\psi(\pi/\chi)}_{ab}\propto\ket{\alpha}_{a}\left\{\ket{\gamma}+\ket{-\gamma}\right\}_{b}+\ket{-\alpha}_{a}\left\{\ket{\gamma}-\ket{-\gamma}\right\}_{b}.
\end{equation}

This is a particular expression for an entangled coherent state: unitarily acting on the subsystem $b$ we can transform it into the more familiar form $\ket{\alpha}_{a}\ket{\gamma}_{b}+\ket{-\alpha}_{a}\ket{-\gamma}_{b}$. In Eq.~(\ref{quasicatstates}), the superpositions of coherent states $\ket{\gamma}_{b}$ and $\ket{-\gamma}_{b}$ are
Schr\"odinger cat states:
\begin{equation}
\label{pariedispari}
\begin{aligned}
&\ket{\gamma}_{b}+\ket{-\gamma}_{b}\propto\sum^{\infty}_{j,0}\frac{\gamma^{2j}}{\sqrt{(2j)!}}\ket{2j}_{b},\\
&\ket{\gamma}_{b}-\ket{-\gamma}_{b}\propto\sum^{\infty}_{j,0}\frac{\gamma^{2j+1}}{\sqrt{(2j+1)!}}\ket{2j+1}_{b}.
\end{aligned}
\end{equation}
These are sometimes called even and odd coherent states. We have shown how the described non-linear interaction, for the particular case of initially prepared coherent states, results in an entangled state. 

\begin{figure}[ht]
\centerline{\psfig{figure=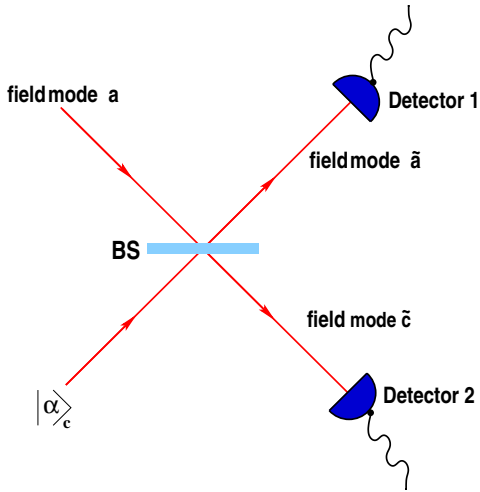,width=6.3cm,height=6.5cm}}
\caption{Scheme to infer the state of the field mode
$a$. It is shown the symbol used for a
photodetector and that for the 50:50 beam splitter (BS). Detector $1$ ($2$) clicks just if the state of
mode $a$ is $\ket{\alpha}_{a}$ ($\ket{-\alpha}_{a}$). With this 
scheme, we can generate an even or an
odd coherent state of mode $b$.} 
\label{detection}
\end{figure}

To prove the entanglement, we have to show the correlation of the modes $a$ and $b$ as we unitarily transform, gradually, from $\ket{\gamma}_{b}$ to $\ket{\gamma}_{b}\pm\ket{-\gamma}_{b}$. However, this involves another non-linear interaction. We thus discuss an indirect procedure. In details, if we can discern where field $\hat{a}$ is, the state of field $\hat{b}$ is projected onto one of $\ket{\gamma}_{b}\pm\ket{-\gamma}_{b}$. To determine the state of field $\hat{a}$ we use a $50:50$ beam splitter (BS) and two photodetectors, as sketched in Fig.~\ref{detection}. After passing through a BS, two coherent input fields $\ket{\alpha},\ket{\beta}$ become:
\begin{equation}
\label{bs}
\hat{B}_{ac}\ket{\alpha}_{a}\ket{\beta}_{c}=\ket{\frac{\alpha+\beta}
{\sqrt{2}}}_{\tilde{a}}\ket{\frac{-\alpha+\beta}{\sqrt{2}}}_{\tilde{c}},
\end{equation}
where $\hat{B}_{ac}\equiv{e}^{\frac{\pi}{4}(\hat{a}^{\dagger}\hat{c}-\hat{a}\hat{c}^{\dagger})}$ is the BS operator,
with $a$ ($\tilde{a}$) and $c$ ($\tilde{c}$) its input (output) modes. For $\beta=\alpha$ we have the following read-out: if the input mode $\hat{a}$ is prepared in $\ket{\alpha}_{a}$, then Detector $1$ will reveal some photons, while Detector $2$ will not. In this case, the field mode $\hat{b}$ will be projected in the even coherent state. In the opposite event, mode $\hat{b}$ will be in the odd coherent state. Of course, there is a possibility to have both the detectors not to click: we do not know in which state mode $\hat{a}$ is and we have to repeat
the procedure untill one of the detectors clicks. 

A possible way to detect the quantum nature of the state generated by the described scheme is schematized as follows: to indirectly infer the coherences in an even or odd coherent state we generate a new entangled coherent state, mixing field mode $\hat{b}$ (which is supposed to be in one of the states in Eq.~(\ref{pariedispari})) and the vacuum of an auxiliary mode in a $50:50$ BS. Calling $\tilde{b},\tilde{c}$ the output modes of the BS, the resulting joint state of radiation is
\begin{equation}
\begin{aligned}
&\hat{\rho}'_{\tilde{b}\tilde{c}}={\cal A}\left\{\pro{\frac{\gamma}{\sqrt 2},\frac{-\gamma}{\sqrt 2}}{\frac{\gamma}{\sqrt 2},\frac{-\gamma}{\sqrt 2}}+\pro{\frac{-\gamma}{\sqrt 2},\frac{\gamma}{\sqrt 2}}{\frac{-\gamma}{\sqrt 2},\frac{\gamma}{\sqrt 2}}\right.\\
&\left.+c\pro{\frac{-\gamma}{\sqrt 2},\frac{\gamma}{\sqrt 2}}{\frac{\gamma}{\sqrt 2},\frac{-\gamma}{\sqrt 2}}+c\pro{\frac{\gamma}{\sqrt 2},\frac{-\gamma}{\sqrt 2}}{\frac{-\gamma}{\sqrt 2},\frac{\gamma}{\sqrt 2}}\right\}_{\tilde{b}\tilde{c}},
\end{aligned}
\end{equation}
where ${\cal A}$ is a normalization constant. The parameter $c$ takes account of the {\it purity} of the generated Schr\"odinger cat state. If it is $c=1$ ($c=-1$), the field mode $\hat{b}$ was in an even (odd) coherent state, while for $c=0$, it was a statistical mixture that has been produced by the non-linear interaction. To discern between the possible values for $c$, we use the criterion for inseparability proposed in~\cite{duan} and we evaluate the function $S=\valmed{(\Delta\hat{u})^2}+\valmed{(\Delta\hat{v})^2}$, with $(\Delta\hat{u})^2$ and $(\Delta\hat{v})^2$ the variances of $\hat{u}=\hat{x}_{\tilde{b}}+\hat{x}_{\tilde{c}}$, $\hat{v}=-\hat{p}_{\tilde{b}}+\hat{p}_{\tilde{c}}$ and $\left\{\hat{x}_{j},\hat{p}_{j}\right\}$ the phase-space quadrature operators for mode $j=\tilde{b},\tilde{c}$~\cite{knightsqueezed}.

\begin{figure}[ht]
\centerline{\psfig{figure=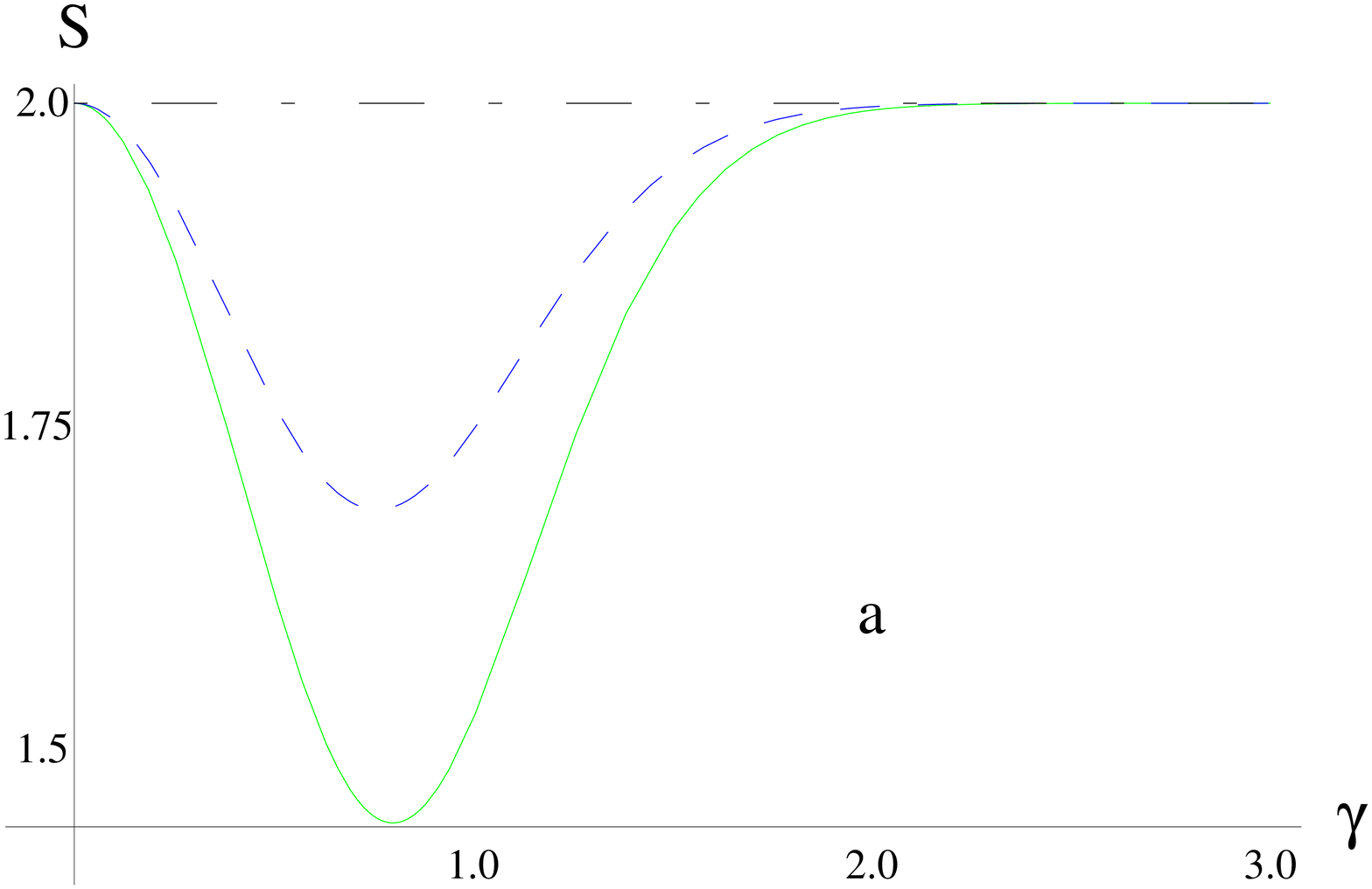,width=6.5cm,height=4.5cm}}
\centerline{\psfig{figure=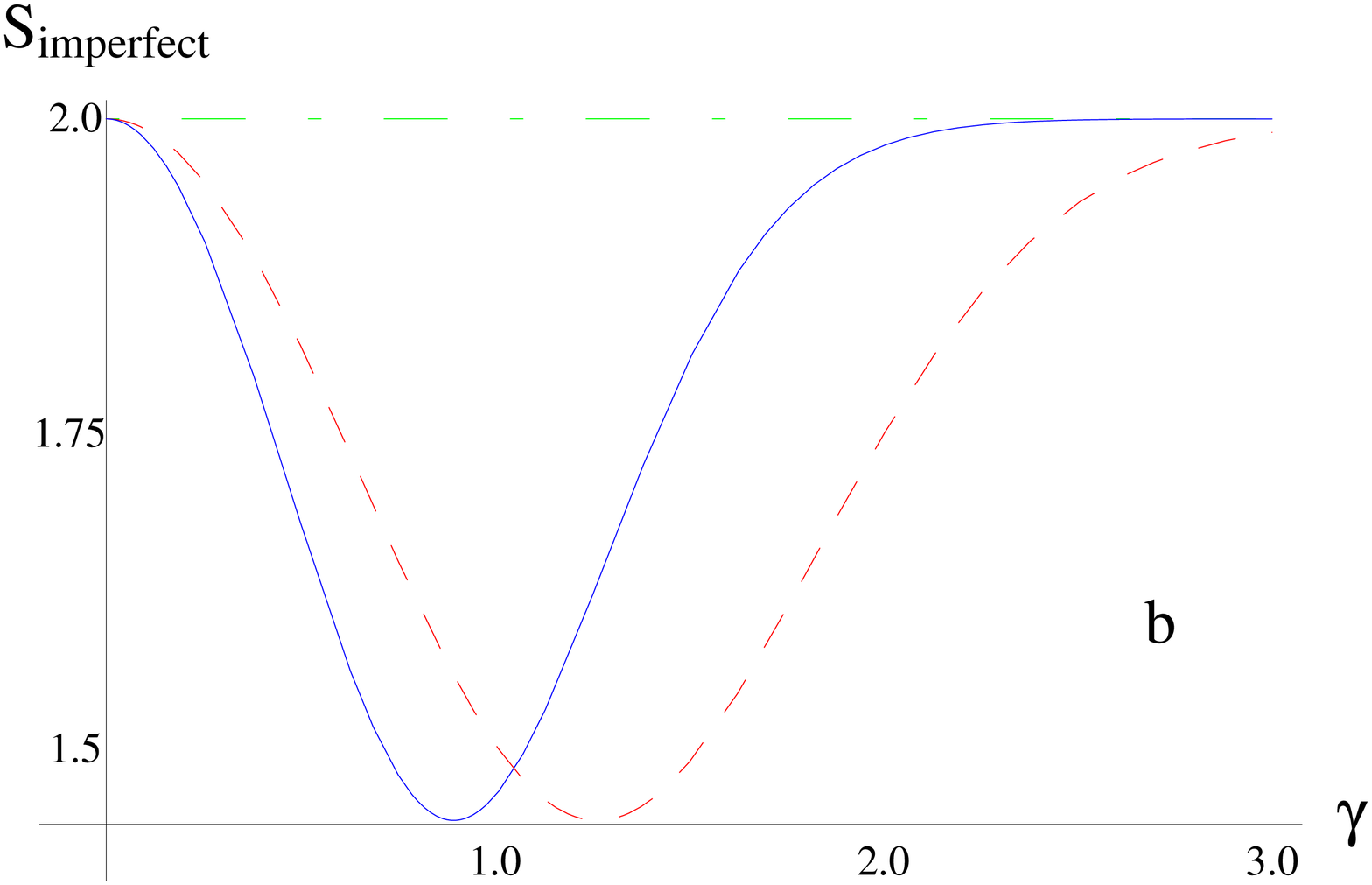,width=6.5cm,height=4.5cm}}
\caption{{\bf(a)} Plot of $S=\valmed{(\Delta{u})^2}+\valmed{(\Delta{v})^2}$ as a function of 
the amplitude $\gamma$. The effect of different values of the 
parameter $c$ in the density matrix $\hat{\rho}'_{\tilde{b},\tilde{c}}$ is studied: the dot-dashed curve is for $c=0$, 
corresponding to the case of a statistical mixture. The dashed curve is for $c=0.5$ while the solid curve represents 
a perfectly generated even coherent state. {\bf (b)} Behaviour of the separability function when imperfection in the 
homodyne detection is considered. Here, the 
dot-dashed curve is for detection efficiency $\eta=0$, the dashed one is for $\eta=0.4$ and the solid curve is for $\eta=
0.8$. For lower $\eta$, the minimum values of $S_{imperfect}$ shift toward higher $\gamma$ values: this is because the 
lower is the efficiency of the homodyne detectors, the more the input state resembles a Gaussian state.}
\label{separabilita}
\end{figure}

According to the sufficient condition for inseparability in~\cite{duan}, if $S\le{2}$, the state of $\tilde{b}$ and $\tilde{c}$ is inseparable. To experimentally evaluate $S$ we need the single quadrature variance ($\valmed{(\Delta\hat{x}_{a,b})^2}$ for example) and correlations as $\valmed{\hat{x}_{a}\hat{x}_{b}}$ or $\valmed{\hat{p}_{a}\hat{p}_{b}}$. All these quantities can be determined via two homodyne detectors, one for each field mode~\cite{knightsqueezed,kimmunro}. A plot of $S$ as a function of the amplitude $\gamma$ in the case of an even coherent state is given in Fig.~\ref{separabilita} ({\bf a}). The bound $S=2$ is violated, for $c=1$ just until $\gamma\simeq2$ (we have $S=1.995$, for $\gamma=2$), revealing the entanglement of the generated state. Our criterion here is only a sufficient condition for entanglement which works fine for a small number of photons. As $\gamma$ grows, this sufficient condition does not give information on entanglement.

It is possible to introduce the homodyne detectors losses modeling an inefficient homodyne detector with a beam splitter BS$_{\eta}$ (transmittivity $\eta$) followed by a perfect homodyne detector~\cite{YurkeStoler}. Each beam splitter BS$_{\eta}$ mixes a mode of the signal to measure with a vacuum state and transmits the signal with probability $\eta$; the amount of reflected input field is a measure of the losses. The quantum efficiency of the detectors is, thus, $\eta$. The calculation of the total variance for the quadratures of modes $\tilde{b},\tilde{c}$ when the detectors have an equal quantum efficiency $\eta$ leads to what is shown in Fig.~\ref{separabilita} ({\bf {b}}). The separability function keeps its functional features even in the case of lossy detection and some similarities with the case of perfect detection are evident, showing the robustness of the scheme. 


\section{Bi-chromatic photon blockade}
\label{photonblockade}

\begin{figure}[ht]
\centerline{\psfig{figure=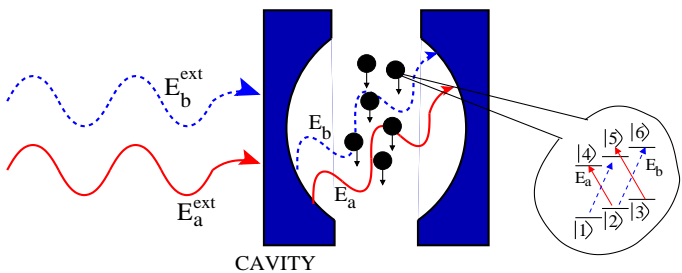,width=9.0cm,height=3.5cm}}
\caption{Set-up for a bi-chromatic photon blockade via large non-linearity. The optical cavity is 
crossed by a low density beam of atoms, each one having the six-level energy scheme suitable for a double-EIT regime. 
This condition is established by two classical driving fields (not shown in the picture) and by two cavity field modes, 
driven on-resonance by two external fields, $E^{ext}_{a}$ and $E^{ext}_{b}$.}
\label{photblock}
\end{figure}

In this Section we give the outlines for a possible quantum control of light offered by the giant non-linearity, in the quantum regime, of the model for double-EIT. In particular, we analize the interaction of a low density group of atoms, with the energy scheme sketched in Fig.~\ref{Petrosyan}, with two modes of an optical cavity. Each cavity mode takes the place of one of the probes in Section~\ref{doubleEIT} and the cavity itself is driven by two weak external beams, each one on-resonance with a relevant cavity mode. The driving fields at frequencies $\omega_{d1},\omega_{d2}$ are then shined on the atoms to obtain the double-EIT regime. The physical system is schematically shown in Fig.~\ref{photblock}.

For the moment, we treat the case in which the density of the atomic beam is so low that the cavity is crossed by a single atom, each time. In this condition, the Hamiltonian of the system {\it atom+cavity+external fields} is $H=H'+\hbar\left\{\omega_{a}\hat{a}^{\dagger}\hat{a}+\omega_{b}\hat{b}^{\dagger}\hat{b}\right\}+\hbar{\cal E}_{pump}\left\{(\hat{a}^{\dagger}+\hat{a})+(\hat{b}^{\dagger}+\hat{b})\right\}$, where $H'$ is the Hamiltonian in Eq.~(\ref{Hamiltonianaquantizzata}) and the third term takes account for the coupling of the cavity with the external fields. The parameter ${\cal E}_{pump}$ is taken the same for both $E^{ext}_{a}$ and $E^{ext}_{b}$. The same arguments used in Section~\ref{appendice} lead to neglect, for the moment, the couplings to the atomic level $\ket{5}$ while, in general, terms for the damping of the cavity modes have to be added. We will consider these two points later. As before, the initial state for the atoms is $\ket{2}$.

We examine what happens if $E^{ext}_{a}$ shines the cavity, exciting the corresponding cavity mode. In this case, the six-level atomic model reduces to a simpler {\sf N} configuration, where the perturbation to the otherwise perfect EIT regime for $E_{a}$ is due to the $\ket{1}\leftrightarrow\ket{5}$ coupling by $E_{a}$ itself. The resulting non-linear self-phase modulation effect, described in Section~\ref{doubleEIT} can not be neglected because the cross phase analog is not active. A mono-chromatic photon blockade effect results, as studied in Ref.~\cite{Imamoglublockade,Rebic,Greentree}. Once a photon in $E^{ext}_{a}$ leaks into the cavity and feeds $E_{a}$, no other external photons are allowed to enter (the photons are {\it blocked}). This is because the self-phase effect gives non-linear features to the atom+cavity mode system: in a dressed-state picture, a second external photon is resonant with no one of the transitions that lead from a singly excited dressed state to a doubly excited one. When the mono-chromatic photon blockade is active, the dressed system atom+cavity mode is trapped between the bare state $\ket{2,0}_{atom,E_{a}}$ and the dark state $\ket{D_{a}}={\cal N}\left\{\Omega_{d}\ket{2,1}-g_{a}\ket{1,0}\right\}_{atom,E_{a}}$ (with a normalization factor {$\cal N$}) and behaves as an effective two-level system. The arguments above can be reformulated when $E^{ext}_{b}$ feeds the cavity.

We now treat the situation in which the cavity has been already fed by one photon (for example from $E^{ext}_{a}$, so that $E_{a}$ is excited) and we consider the effects of the interaction with a photon of different {\it colour} (i.e. with a photon from $E^{ext}_{b}$). The initial state is $\ket{2,0,0}_{atom,E_{a},E_{b}}$, that is the state without excitation. The absorption of the first photon by the cavity takes the system to the dressed state $\ket{D_{a}}$. To see if a photon blockade with respect to incoming photons of frequency $\omega_{b}$ is possible, we diagonalize the interaction Hamiltonian in the subspace spanned by states as $\ket{atom}\otimes\ket{E_a}\otimes\ket{E_b}$ that have two excitations.  

\begin{figure}[ht]
\centerline{\psfig{figure=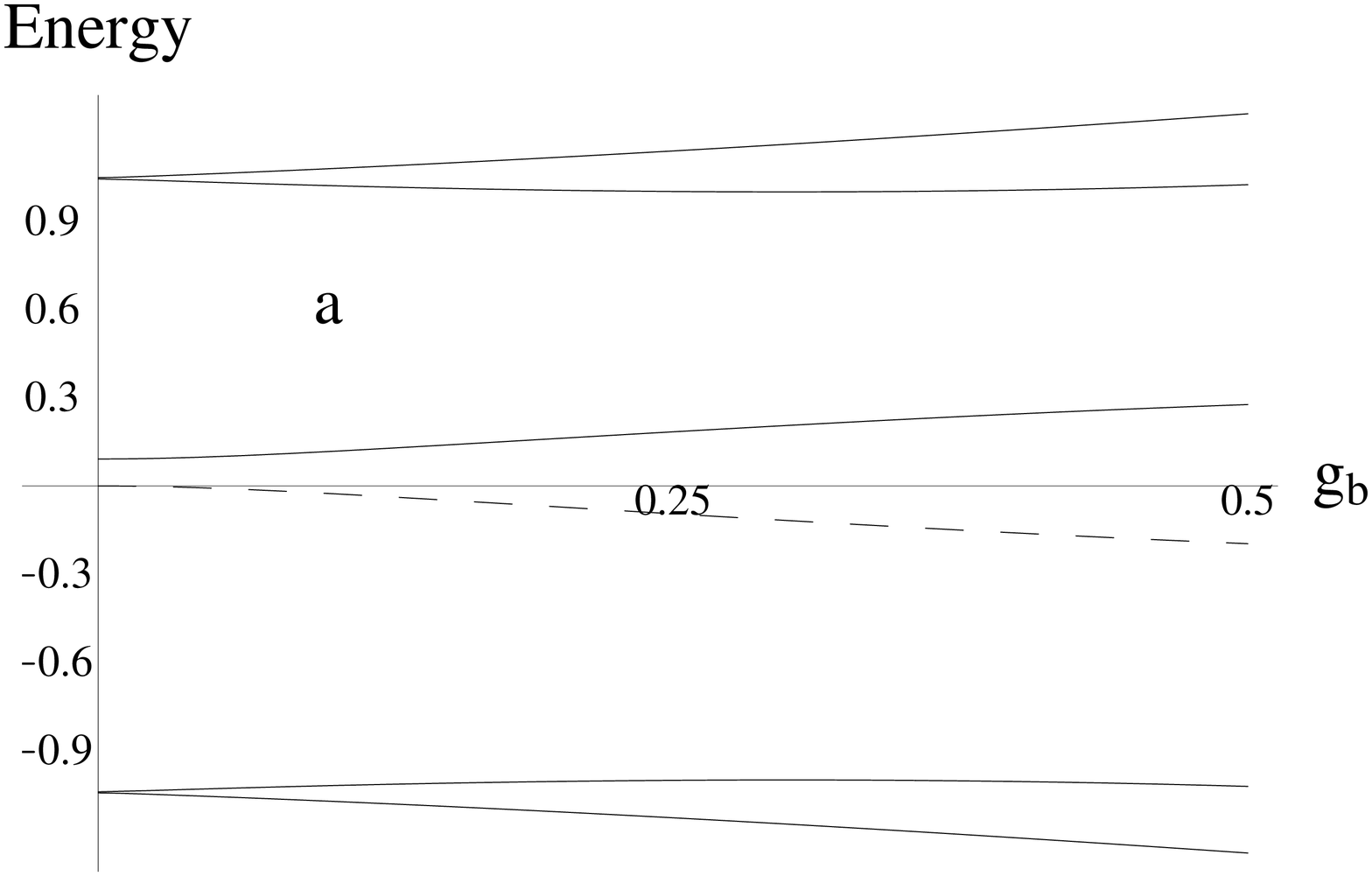,width=8.0cm,height=5.5cm}}
\centerline{\psfig{figure=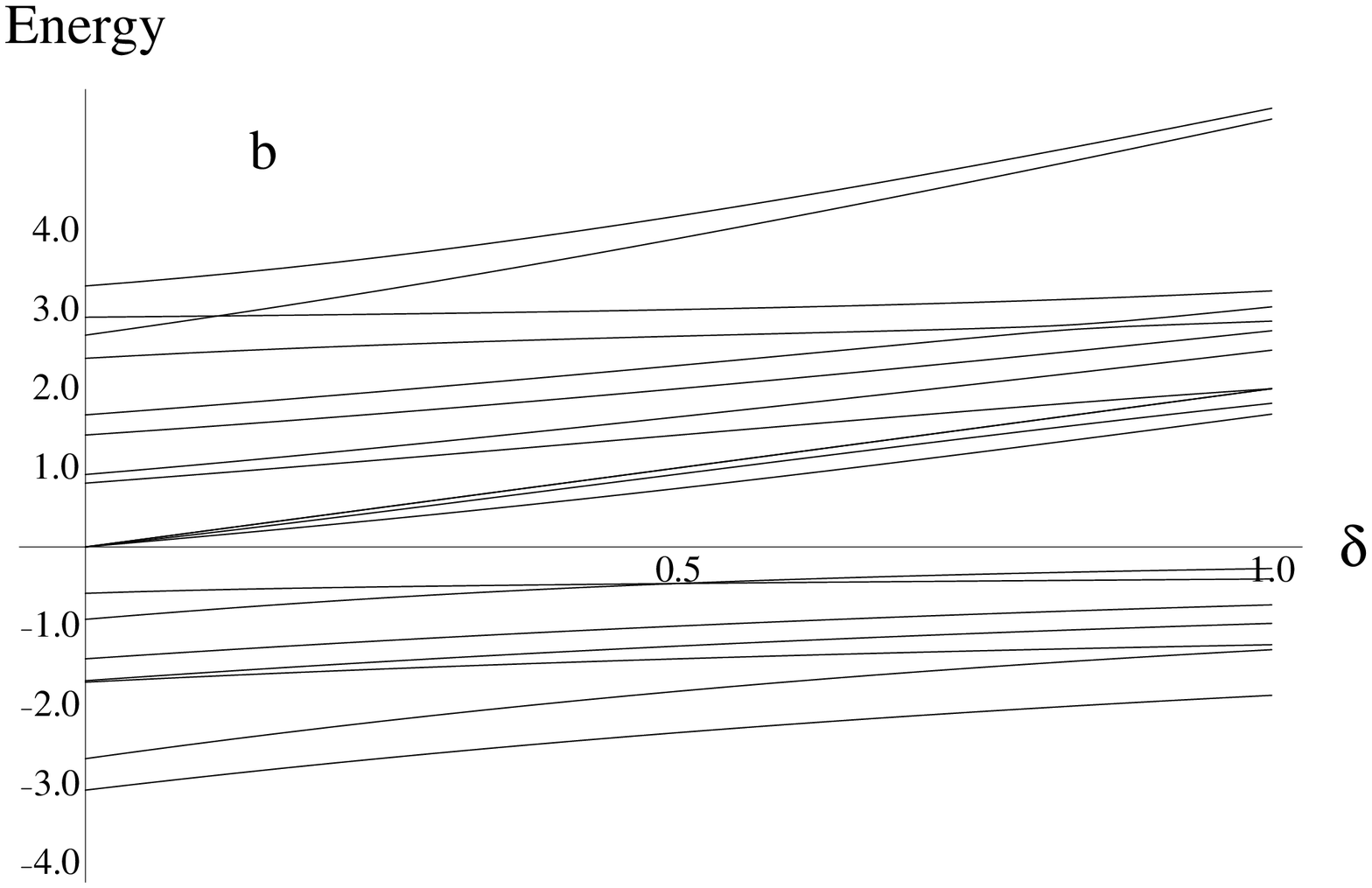,width=8.0cm,height=5.5cm}}
\caption{{\bf (a)} Eigenenergy for the doubly excited manifold of the system in Fig.~\ref{photblock}, when 
the excitations are shared between both the cavity modes (energies and $g_{b}$ in units of $\Omega_{d}$). $g_{a}/\Omega_{d}=0.3$, 
$\gamma/\Omega_{d}=0.01$ and $\Delta/\Omega_{d}=0.1$. The photon blockade effect with respect to a photon of {\it colour} $\omega_{b}$ 
is evident if $g_{b}\neq0$. {\bf (b)} Restoration of photon blockade for a non-zero value of the 
single photon detuning $\delta$ when two atoms are present in the cavity. The introduction of $\delta$ drives the 
eigenstates of the doubly excited manifold out of resonance from the cavity field. No resonant 
transition is now found, showing the restored effectiveness of a blockade effect. The same 
parameters are used as in {\bf (a)}.}
\label{h11con1atomo}
\end{figure}

The interaction Hamiltonian, taking explicitly into account the couplings to $\ket{5}$, is closed within $\left\{\ket{2,1,1},\ket{4,0,1},\ket{1,0,1},\ket{6,1,0},\ket{3,1,0},\ket{5,0,0}\right\}$. We take $\Delta\ll\Omega_{d}$ and $g_{a}\neq{g}_{b}$. As a function of the Rabi frequency $g_{b}$, a typical plot of the eigenenergies is shown in Fig.~\ref{h11con1atomo} ({\bf a}). A photon blockade effect is evident: to feed the cavity, a photon of frequency $\omega_{b}$ should find an eigenenergy exactly equal to zero in the plot (the energy scale is referred to $\hbar\omega_{a}+\hbar\omega_{b}$, so that the plot shows only the non-linear part of the eigenspectrum). For $g_{b}\neq0$ there is not such a possibility: the energies of the six dressed eigenstates are all non-zero (the flatness of the dashed energy curve is just an effect of the large scale in the energy axis of the plot). The incoming photon can not leak into the cavity. Analogously, a photon of $E_{a}$ can not penetrate into the cavity if a $E^{ext}_{b}$ photon is already there. 

The effects of the cavity dissipation and the atomic decay can be considered adding the term $-i\hbar\Gamma(\hat{a}^{\dagger}\hat{a}+\hat{b}^{\dagger}\hat{b})-i\hbar\gamma\sum^{6}_{j=4}\pro{j}{j}$, where $\Gamma$ is the decay rate of the cavity and $\gamma$ is the atomic decay rate as in Section~\ref{doubleEIT}. If we consider $\Gamma,\gamma\ll\Omega_{d}$, however, their effects are to give a small linewidth to each eigenvalue, in any case insufficient to create suitable conditions for transitions to the doubly excited manifold. The external driving term induces transitions between the two levels 
in which the system has been effectively reduced. From the optical Bloch equations for a field interacting with a two-level atom (with a decay rate small compared to the Rabi frequency of the interaction) in stationary conditions, the population of the excited state of the atom is expected to reach $\sigma^{stat}_{kl}\simeq1/2$ (the subscripts $k,l$ state the number of excitations in the ground and excited states). In our case, it is easy to prove that $\modul{{\cal E}_{pump}\bra{down}\hat{a}\ket{up}}$ plays the role of a Rabi frequency for the interaction between the external field and the cavity+atom system, whose ground and excited states are labeled as $\ket{down}$ and $\ket{up}$. For the coupling between $\ket{down}\equiv\ket{2,0,0}$ and $\ket{up}\equiv\ket{D_{a}}\otimes\ket{0}_{E_{b}}$, with ${\cal E}_{pump}\simeq{0}.1\Omega_{d}$ (weak pump regime) and the parameters in Fig.~\ref{h11con1atomo}, we get $\sigma^{stat}_{01}\simeq0.497$. If the same calculation is performed with respect to the coupling between $\ket{down}\equiv\ket{D_{a}}\otimes\ket{0}_{E_{b}}$ and $\ket{D_{ab}}$, which is the state corresponding to the dashed curve in Fig.~\ref{h11con1atomo} ({\bf a}), we get $\sigma^{stat}_{12}\simeq0.091\ll{1/2}$. 
In the above discussion, we assumed the single-atom interaction. Let us consider a case in which a number of atoms are present in the cavity. This could happen because, experimentally, we do not have a perfect control of the number of atoms that cross the cavity. It has been shown in~\cite{Imamoglublockade} indeed that having more than a single atom in the cavity has the effect to introduce a large number of other energy levels that accumulate near the zero energy axis. This makes the photon blockade less effective. If we analyze our specific system, we see that the Hamiltonian itself acquires a {\it collective operator} structure. In general, atomic operators such as $\pro{i}{j}$ are replaced by $\sum^{N}_{k,1}\ket{i}_{k}\bra{j}$, with $k$ labeling the $N$ particles inside the cavity. The injection of a photon into the cavity couples $\ket{\underline{2}}_{atoms}\otimes\ket{0}_{E_{a}}$ to$\ket{\underline{D}_{a}}$ where we introduced the collective states: 
\begin{equation}
\label{dickestates}
\begin{aligned}
&\ket{\underline{2}}_{atoms}\otimes\ket{0}_{E_{a}}\equiv\ket{2..2}_{atom1..atomN}\otimes\ket{0}_{E_{a}},\\
&\ket{\underline{D}_{a}}=\frac{\Omega_{d}\ket{\underline{2}}_{atoms}\otimes\ket{1}_{E_{a}}-g_{a}\ket{\underline{1}}_{atoms}\otimes\ket{0}_{E_{a}}}{\sqrt{\Omega^2+N{g}^2}}, 
\end{aligned}
\end{equation}
with
\begin{equation}
\nonumber
\ket{\underline{1}}_{}\equiv\frac{1}{N}\left\{\ket{122..2}+\ket{212..2}+..+\ket{222..1}\right\}_{atom1..atomN}
\end{equation}
 being a singly excited symmetric Dicke state. To check the possibility for a photon blockade we have to seek the eigenvalues in the doubly excited manifold.
Let us consider the simple case of $N=2$. The manifold with a single excitation is now composed of five bare states while they were three for the single atom case and there is an energy equal to $\hbar\omega_{a}$ ($\hbar\omega_{b}$) if a photon from $E^{ext}_{a}$ ($E^{ext}_{a}$) has excited the cavity field. Transitions to the singly excited manifold are possible. If we consider the doubly-excited subspace, we find eigenenergies suitable for a resonant transition that ruins the blockade effect. This behaviour can be verified for the cases of $N=3,4$ and we conjecture that this feature is present for an arbitrary $N$. To bypass this problem, we introduce a single photon detuning in the transitions $\ket{2}\leftrightarrow\ket{4}$, $\ket{2}\leftrightarrow\ket{6}$, while keeping the two-photon resonance. This implies the introduction of $\hbar\delta\sum_{j}(\ket{4}_{j}\bra{4}+\ket{6}_{j}\bra{6})$ in the Hamiltonian. If $\delta>\gamma$ these terms take account of the detuning of the dressed states off the cavity resonance: it shifts the energies in the doubly excited manifold and restores the blockade. The absence of resonant energies for $N=2$, $\delta\neq0$ and $\delta\gg\gamma$ is shown in Fig.~\ref{h11con1atomo} {\bf (b)}. We have extended this investigation up to the case of $N=4$. 
These results, obtained for values of the involved parameters achievable by current technology~\cite{mio}, are in agreement with the analysis in~\cite{Greentree}. 

\begin{figure}[ht]
\centerline{\psfig{figure=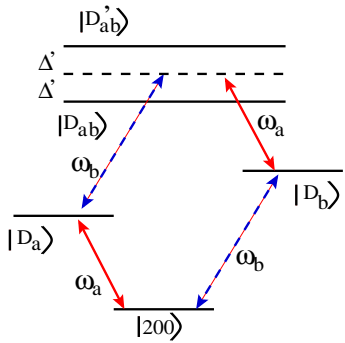,width=4.5cm,height=4.5cm}}
\caption{Effective five-level model for the atom+bi-modal cavity system, when the doubly excited eigenstates 
($\ket{D_{ab}}$ and $\ket{D'_{ab}}$) nearest to a resonant coupling with the single-quantum manifolds are included and we assume $\omega_{a}<\omega_{b}$. 
The detuning of $\ket{D_{ab}}$, $\ket{D'_{ab}}$ from $\omega_{a}+\omega_{b}$ is symmetric for $g_{a},g_{b}=g
\simeq0.5\Omega_{d}$.}
\label{interferenza}
\end{figure}

We now introduce a quantum interference effect that explains the inhibition of transitions to highly excited states (the low values for $\sigma^{stat}_{kl}$ can not be explained just by the detuning from resonance~\cite{Imamoglublockade}). For simplicity, we consider the case of a single atom inside the cavity. If the external fields shine together the cavity, they simultaneously try to excite the cavity field but, as long as one photon is fed into the cavity, there is no possibility for a second one to leak. We assume $\omega_{a}<\omega_{b}$ and we include the two eigenstates, $\ket{D_{ab}}$ and $\ket{D'_{ab}}$, that in Fig.~\ref{h11con1atomo} ({\bf a}) are the nearest to the zero energy axis. This permits to estimate the blockade effect from a different perspective. The resulting effective five-level model is in Fig.~\ref{interferenza}. Taking a value of $g_{a},{g}_{b}\sim0.5\Omega_{d}$, the splitting of $\ket{D_{ab}}$ and $\ket{D'_{ab}}$ from the resonance becomes symmetric (as can be seen in Fig.~\ref{h11con1atomo} {\bf (a)}, even for $g_{b}=0.5\Omega_{d}$, $g_{a}=0.3\Omega_{d}$) and we take it to be $\Delta'$.

\begin{figure} [ht]
\centerline{\psfig{figure=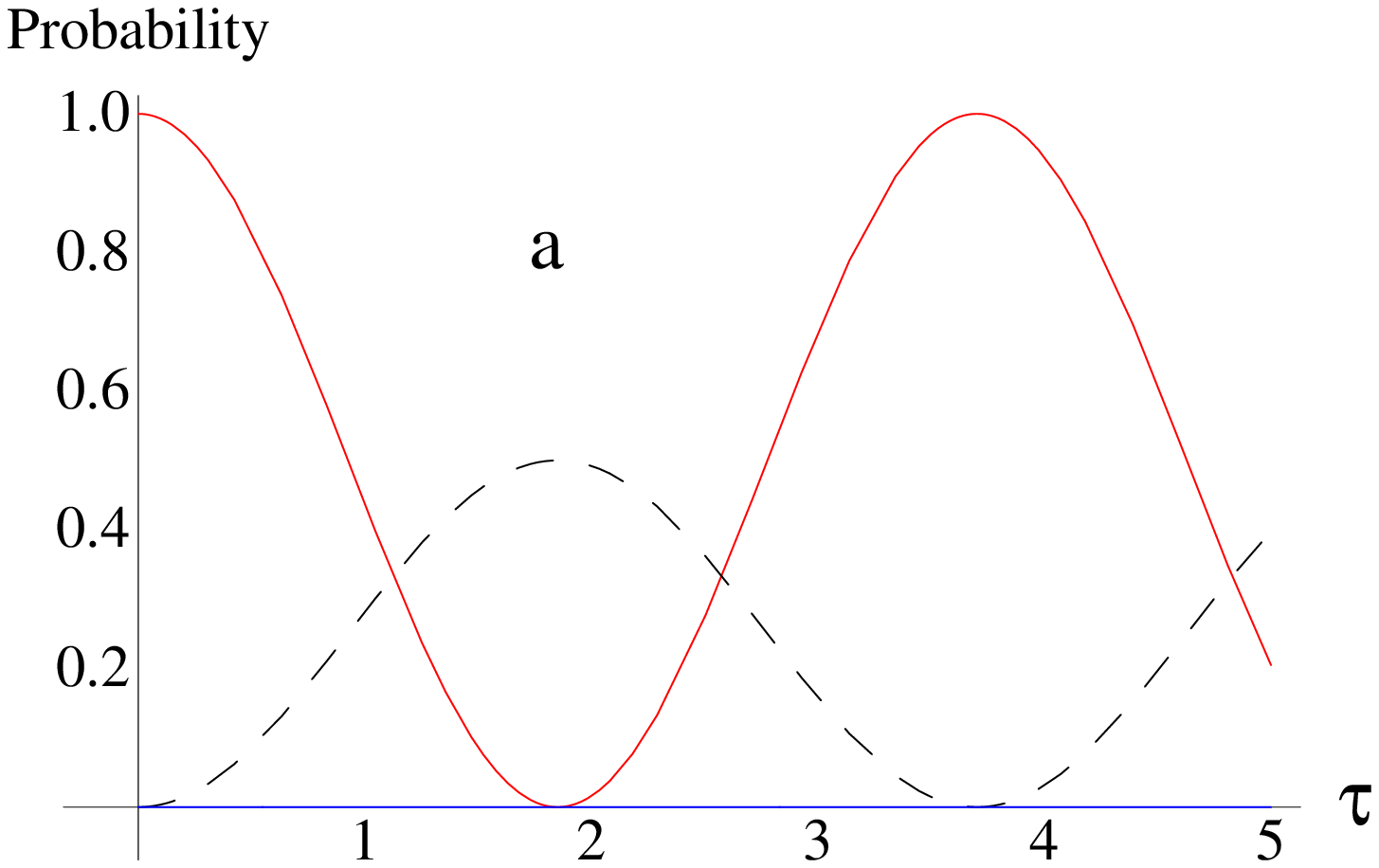,width=6cm,height=4cm}}
\centerline{\psfig{figure=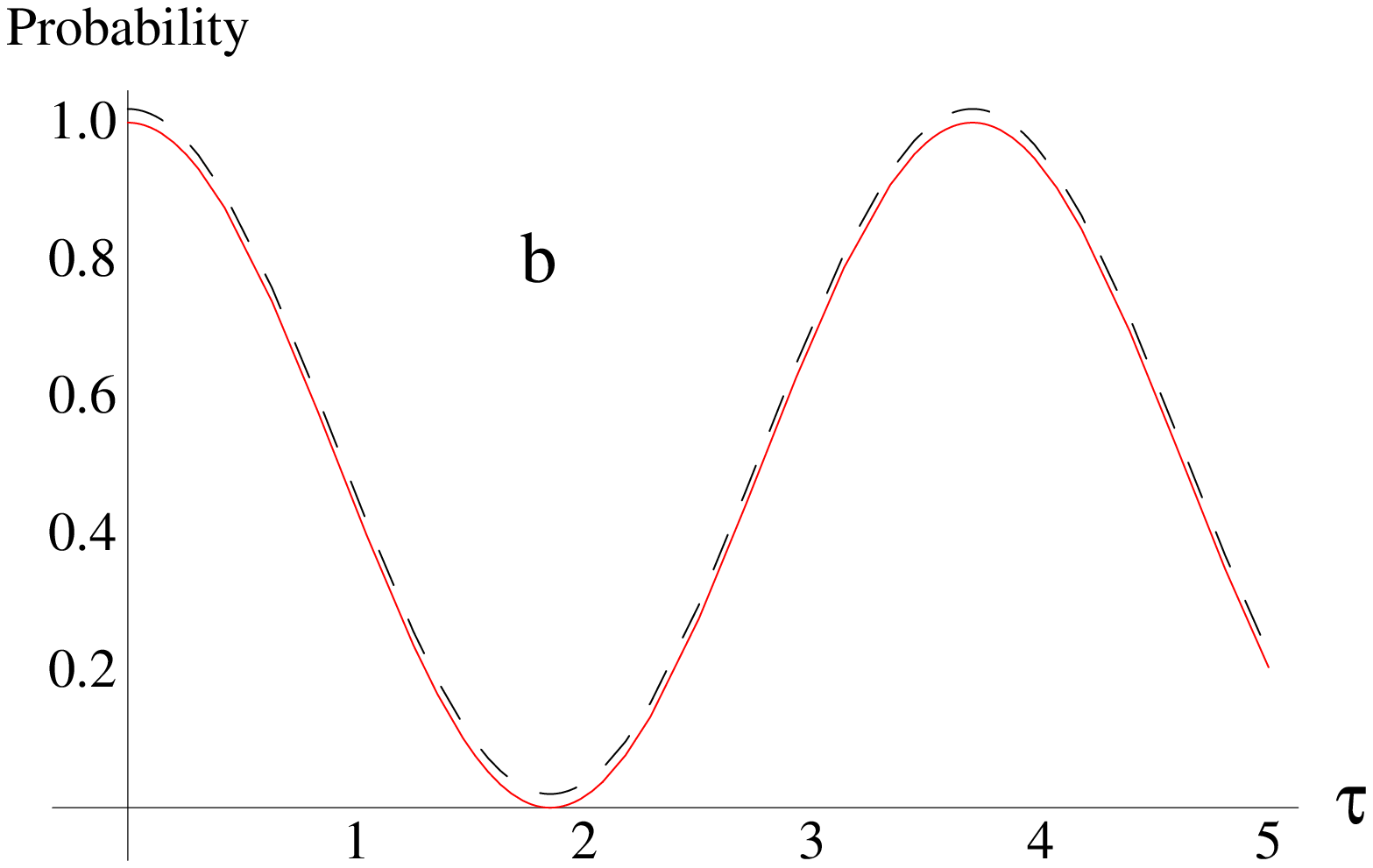,width=6cm,height=4cm}}
\caption{{\bf (a)} Plot of the probabilities $\modul{C_{1}}^2$ (solid line), $\modul{C_{2}}^2$ 
and $\modul{C_{3}}^2$ (dashed line). The latter two are completely superimposed. $\tau$ is a 
rescaled time: $\tau=t/\Omega_{R}$, with $\Omega_{R}$ the effective Rabi frequency of the 
oscillations. {\bf (b)} Comparison between the analytical (solid line: the same as in {\bf (a)}) 
and the numerical (dashed line) solution for $\modul{C_{1}}^2$. The small offset of the 
numerical solution has been added to make the curve visible: the matching with the analytic 
solution is perfect.}
\label{results}
\end{figure}

The effective Hamiltonian, in the basis composed of $\left\{\ket{200}, \ket{D_{a}},\ket{D_{b}},\ket{D_{ab}},\ket{D'_{ab}}\right\}$ can be easily written for the couplings shown in Fig.~\ref{interferenza} and its application to a state of the form
\begin{equation}
\label{statointerferenza}
\ket{\eta(t)}=C_{1}\ket{200}+C_{2}\ket{D_{a}}+C_{3}\ket{D_{b}}+C_{4}\ket{D_{ab}}+C_{5}\ket{D'_{ab}},
\end{equation}
leads to a Schr\"odinger equation that can be recast in a set of five differential equations for the coefficients $C_{j}$ $(j=1,..,5)$. We expect that the approximation of two-level system, consequence of the photon blockade, has to be good, so that the populations of states $\ket{D_{ab}},\ket{D'_{ab}}$ should remain very small. Numerical integration of these equations confirms our expectations and we get $\modul{C_{4}}^2=\modul{C_{5}}^2\leq10^{-2}$. We thus neglect these terms and we find a simple analytical solution that results in Rabi oscillations of the populations of states $\ket{200}$, $\ket{D_{a}}$, $\ket{D_{b}}$ with the frequency $\Omega_{R}=\Omega_{d}{\cal E}_{pump}/\sqrt{\Omega^2_{d}+g^2}$,
that matches the analogous parameter from $\sigma^{stat}_{kl}$. The results are shown in Figs.~\ref{results}. 
The amplitudes of oscillation for $\modul{C_{2}}^2$ and $\modul{C_{3}}^2$ are $1/2$, and they oscillate in phase (absolutely indistinguishable in the plot). This is because we do not know which one of them could have taken the excitation from outside. The probability of excitation is thus $1/2$ for both the fields. At this time, the state of the system in an equally weighted superposition of $\ket{D_{a}}$ and $\ket{D_{b}}$. The calculation of $C_{2}$ and $C_{3}$ leads to: 
\begin{equation}
\label{inter}
\begin{aligned}
&\frac{1}{\sqrt 2}\left\{\cos{\vartheta_{a}}\ket{10}_{ab}+\cos{\vartheta_{b}}\ket{01}_{ab}\right\}\otimes\ket{2}_{atom}\\
&-\frac{1}{\sqrt 2}\left\{\sin{\vartheta_{a}}\ket{1}_{atom}+\sin{\vartheta_{b}}\ket{3}_{atom}\right\}\otimes\ket{00}_{ab}
\end{aligned}
\end{equation} 
with $\cos{\vartheta_{a,b}}=\Omega_{d}/\sqrt{\Omega^2_{d}+g^2_{a,b}}$. If we measure the state of the atom, as it exits from the cavity, and we find $\ket{2}$, the cavity modes are projected onto an entangled state with adjustable coefficients. Even if we have just roughly outlined the problem, this certainly deserves a further analysis. This result seems to be promising in the perspective of quantum state engineering, in particular because no single-atom addressing is required.

\section{Conclusions}
\label{conclusions}
The main result of this work is the discussion of the quantized picture of a model for
double-EIT~\cite{PetrosyanKurizki}. Our approach is based on a full
Hamiltonian method that simplifies the problem of a many-level atomic system interacting with some electromagnetic fields. The results obtained have been confirmed by a dressed state approach. Here, we have shown the cross-phase modulation of two interacting fields. We suggested the Pr:YSO crystal as a candidate to embody the model. The success of the quantization step has led us to investigate two particular problems in the context of quantum state engineering: the generation of entangled coherent states of light and a bi-chromatic photon blockade in CQED. In the former case, a scheme for the inference of the non-classicality of the generated state has been briefly discussed, including the effects of losses by the detection apparatus. 

For the photon blockade, we discussed the main features for the control of the population of two cavity field modes. Considering a flux of atoms crossing a bi-modal cavity, we have shown the effectiveness of the blockade effect even when more than a single atom is present in the cavity. With respect to a solid state system (as Pr:YSO) placed into the cavity, the examined set-up is more suitable for the realization of photon blockade because it overcomes a series of severe restrictions that in a multi-atom system are imposed because of the high-dispersion limit~\cite{Imamoglublockade}. 

In conclusion we have outlined specific applications of a particular kind of large and efficient non-linearity, namely the double-EIT regime, in the context of controlling a quantum system.


\section*{Acknowledgments}

This work is supported by the UK Engineering and Physical Science Research Council through GR/R33304. B.H. acknowledges the financial support from the  Korean Ministry of Science and Technology through the Creative Research Initiative Program. M.P. acknowledges the International Research Centre for Experimental Physics.


\begin{thebibliography}{99}
\bibitem{mio} $\Omega_{d}\sim10$ MHz, $\Gamma\sim10^5$ Hz and $g_{a,b}\sim1$ MHz are reasonable values and, for the $5S_{1/2}\rightarrow{5}P_{1/2}$ in $^{87}$Rb, we can take $\gamma\sim10^5$ Hz, $\delta\sim1$ MHz.
\end{thebibliography}
\end{document}